# Experimental evidence for hotspot and phase-slip mechanisms of voltage switching in ultra-thin YBa$_2$Cu$_3$O$_{7-x}$ nanowires

M. Lyatti[1,2,3,*], M. A. Wolff[2], A. Savenko[4,5], M. Kruth[6], S. Ferrari[2], U. Poppe[7],

W. Pernice[2], R. E. Dunin-Borkowski[3], C. Schuck[2]

[1] Kotelnikov IRE RAS, 125009 Moscow, Russia
[2] Institute of Physics, University of Münster, 48149 Münster, Germany
[3] PGI-5, Forschungszentrum Jülich, 52425 Jülich, Germany
[4] FEI Electron Optics BV, 5651 GG Eindhoven, Netherlands
[5] ZEA-3, Forschungszentrum Jülich, 52425 Jülich, Germany
[6] ER-C 2, Forschungszentrum Jülich, 52425 Jülich, Germany
[7] CEOS GmbH, Englerstr. 28, 69126 Heidelberg, Germany

We have fabricated ultra-thin YBa$_2$Cu$_3$O$_{7-x}$ nanowires with a high critical current density and studied their voltage switching behavior in the 4.2 - 90 K temperature range. A comparison of our experimental data with theoretical models indicates that, depending on the temperature and nanowire cross section, voltage switching originates from two different mechanisms: hotspot-assisted suppression of the edge barrier by the transport current and the appearance of phase-slip lines in the nanowire. Our observation of hotspot-assisted voltage switching is in good quantitative agreement with predictions based on the Aslamazov-Larkin model for an edge barrier in a wide superconducting bridge.

## I. INTRODUCTION

Over the last decade, superconducting nanowires have attracted attention because of their promising applications in quantum sensing and computing [1-3]. Abrupt voltage switching is a characteristic feature of superconducting nanowires and is used to investigate superconductivity in low-dimensional structures, as well as for practical applications. Voltage switching is observed in both low-temperature (low-$T_c$) and high-temperature (high-$T_c$) current-biased superconducting bridges [4-16]. Voltage switching in conventional low-$T_c$ superconducting bridges is well understood [4-7,16-18]. However, there is no consensus for explaining the origin of voltage switching in high-$T_c$ cuprate superconductors. Several mechanisms have been considered to explain discontinuities in the current-voltage (*IV*) characteristics of high-$T_c$ superconducting bridges, including flux-flow instabilities [8,13-15], a phase-slip process [11], hotspot effects [8,12], and fluctuating charge stripe domains [10]. In wide and thick bridges, all of these mechanisms can coexist within the same current range, which complicates the analysis of experimental data. However, the identification of voltage switching mechanisms is possible in superconducting wires whose dimensions approach the characteristic length scales of the superconducting state. As a result of the short coherence length and small magnetic penetration depth of high-$T_c$ cuprate superconductors, suitable wire widths and thicknesses are in the range of a few (tens of) nanometers. The fabrication of such thin and narrow high-$T_c$ nanowires with homogeneous superconducting properties is challenging, both given the material's growth mode [19] and given the limitations of current thin-film patterning technology [20,21].

---

[*] matvey_l@mail.ru

Here, we realize ultra-thin $YBa_2Cu_3O_{7-x}$ (YBCO) films with high critical current densities and smooth surfaces [22] and use focused ion beam (FIB) milling to fabricate YBCO nanowires from them with widths down to 30 nm. We observe voltage switching in current-biased YBCO nanowires and show that it originates from two different mechanisms: hotspot-assisted suppression of the edge barrier and the appearance of phase-slip lines.

## II. YBCO NANOWIRE FABRICATION

YBCO films were deposited on $TiO_2$-terminated (100) $SrTiO_3$ substrates using dc sputtering at a high oxygen pressure of $p(O_2)$ = 3.4 mbar [23]. The temperature of the substrate heater was 935ºC during sputtering, allowing for the simultaneous realization of a good film crystallinity, resulting in a high critical current density, a high critical temperature, and a narrow transition width, and a smooth film surface with a low density of precipitates. After YBCO film deposition, the substrate temperature was ramped down to 500 ºC and the film was annealed in $O_2$ (800 mbar) for 30 min at this temperature. The substrate was then cooled to 50 ºC and the YBCO film was covered *in situ* by a 10-15 nm thick amorphous YBCO layer. Following amorphous YBCO layer deposition, the substrate temperature was ramped up to 500 ºC and the YBCO film was annealed in $O_2$ (800 mbar) for a second time. The amorphous YBCO is not superconducting, but serves as a protective layer for the ultra-thin superconducting YBCO film during the lithography process. The room-temperature resistivity of the amorphous YBCO layer is 700 $\Omega$ cm, which is six orders of magnitude higher than the resistivity of the superconducting YBCO film and therefore has a negligible influence on the electrical measurements that are discussed below [22]. Further details about the ultra-thin YBCO films used in this work are reported elsewhere [22]. After film deposition, 100-nm-thick Au contact pads were fabricated *ex situ* using room-temperature dc magnetron sputtering with a shadow mask. In order to achieve a better electrical contact, the amorphous YBCO layer was removed from the contact pad areas before Au deposition.

A two-step process was used to pattern narrow nanowires from the ultra-thin YBCO films. In the first step, microbridges were defined in the photoresist (PMMA) using optical UV contact lithography. Pattern transfer to the YBCO layer was achieved using chemical etching in a Br-ethanol solution for films that were thicker than 10 nm and Ar ion beam etching for thinner films. In the second step, FIB milling was used to pattern nanowires across the microbridges, as shown in Fig. 1. In order to protect the ultra-thin YBCO film during FIB milling, it was covered with a 20-nm-thick PMMA resist layer and then with a 100-nm-thick Au layer deposited by dc magnetron sputtering at room temperature before the patterning process. The Au layer also serves as a conducting layer, in order to avoid charging during imaging using scanning electron microscopy (SEM) or focused ion beam microscopy. The intermediate PMMA layer was introduced in order to allow the Au layer to be removed after patterning by lift-off.

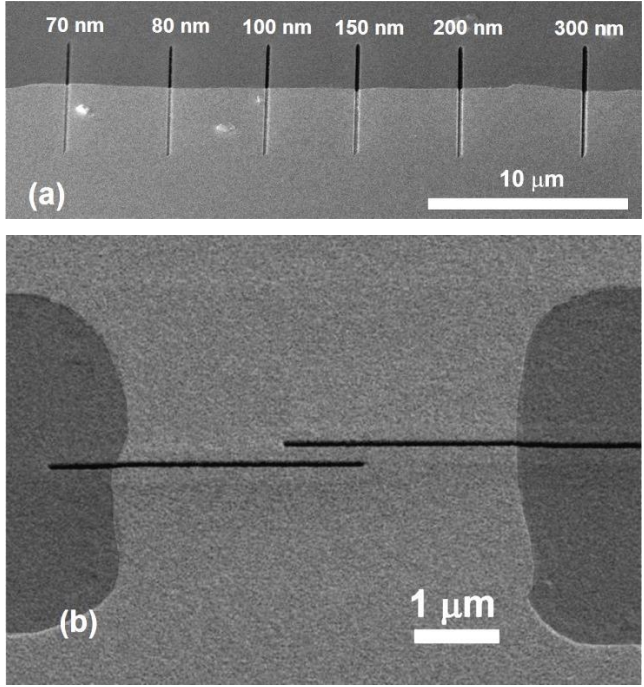

FIG. 1. (a) False-color SEM image of a series of test cuts made using increasing FIB milling times from left to right; (b) false-color SEM image of a nanowire placed across a 5-µm-wide YBCO microbridge. The numbers above the cuts in (a) are the expected milling depths calculated relative to Si. Light grey, dark grey and black correspond to YBCO and SrTiO$_3$ under the Au layer and bare insulating SrTiO$_3$, respectively.

Special attention was paid to minimize nanowire heating during FIB milling. The milling was performed in Helios NanoLab 600i and Helios NanoLab 460F1 instruments (FEI Company) using Ga+ ion beams with accelerating voltages of 30 kV and minimum currents of 7.7 pA. For each sample, we calibrated the minimum milling time by performing a series of test cuts, which were inspected using SEM. An SEM image of the test cuts is shown in Fig. 1a. On increasing the milling time, the region of the cut in the image was observed to change from grey to black, corresponding to the conducting material and insulating SrTiO$_3$ substrate, respectively. The right-most cut in Fig. 1a corresponds to the minimum milling time required to reach the underlying SrTiO$_3$ substrate. In order to further reduce the milling time (and hence heating), the nanowires were patterned using two short cuts across the microbridge, as shown in Fig. 1b. After FIB milling, the Au layer was removed without damaging the YBCO nanowires by immersing the sample in acetone.

### III. ELECTRICAL CHARACTERIZATION OF YBCO NANOWIRES

The influence of the fabrication process on the superconducting properties of the YBCO nanowire was evaluated by comparing electrical transport measurements through the microbridge (before FIB-milling) with similar measurements performed on the nanowire (after FIB-milling).

A battery-operated current source was used for biasing the bridges. The voltage across the bridges was amplified using a battery-operated low-noise amplifier with a frequency bandwidth of 100 kHz. The bias current was modulated with an amplitude of 10 nA at a frequency of 30 kHz and a lock-in amplifier was used to measure the temperature dependence of the normal-state resistance $R_n$. Such ac biasing has minimal influence on measurement-induced alterations of the critical temperature and superconducting transition width.

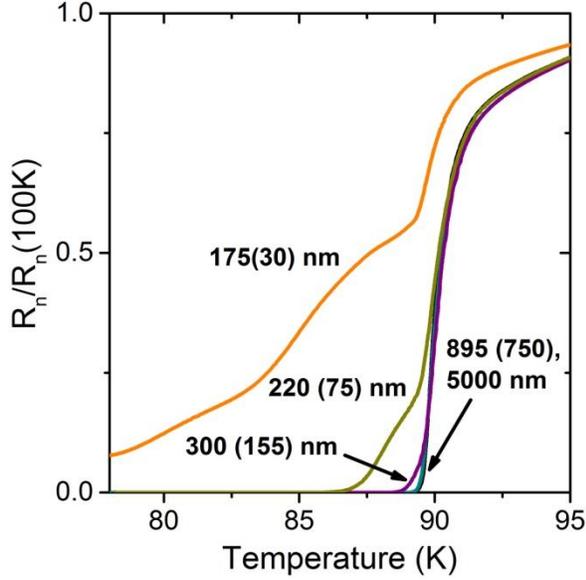

FIG. 2. Temperature dependences of the normal-state resistance of 20 u.c. thick YBCO bridges N1-N5 with various widths. The numbers near the curves are the measured and effective (in brackets) nanowire widths.

Reference $R_n(T)$ curve for a 5-μm-wide and 20 unit cell (u.c.) thick YBCO microbridge N1 is shown in Fig. 2 with the index 5000 nm. Here, 1 u.c. corresponds to a 1.168 nm length in the $c$-axis direction. The 5-μm-wide microbridge N1 shows an $R_n(300K)/R_n(100K)$ ratio of 3.2, a midpoint critical temperature of 90 K, a 90-10% transition width of 1.1 K and a critical current density of 9.7 MA/cm$^2$ at a temperature $T$ of 78 K.

We compared this reference measurement with 1-2 μm long nanowires of various widths below 1 μm produced by FIB milling across a microbridge, as shown in Fig. 1b. The resulting $R_n(T)$ curves, which are shown in Fig. 2, demonstrate that the nanowire N2 with a measured width of 895 nm has a similar critical temperature and superconducting transition width to the reference values for a 5-μm-wide microbridge N1. A reduction in nanowire N3 width to 300 nm results in a slight broadening of the superconducting transition, while the midpoint critical temperature remains unchanged. For narrower nanowires N4 and N5 of width 220 and 175 nm, respectively, significant broadening of the superconducting transition is observed.

In order to assess the influence of the patterning process on the superconducting properties of the nanowires, we measured the room-temperature conductance and *IV* characteristics and extracted the critical currents for different wire widths. The critical currents were determined from the measured *IV* curves with a voltage threshold of 10 μV. In a first series of four-wire measurements on 20 u.c. thick nanowires, the critical currents of 175-895 nm wide nanowires N2-N5 were determined at $T = 78$ K. Fig. 3 shows a linear scaling of both the critical current and the room-temperature conductance with nanowire width. However, a linear fit to the critical current (conductance) data reveals a zero crossing for wires of width $W_d = 162\pm24$ nm ($145\pm7$ nm). Similar behavior is observed in a second series of two-wire measurements of 7 u.c. thick wires, for which critical current measurements performed on 175-405 nm wide nanowires at $T = 4.2$ K are also shown in Fig. 3a. In this case, we find that $W_d = 139\pm8$ nm, in good agreement with the values extracted from measurements on thicker films. We interpret this value as the width of the wire that was damaged during the fabrication process, e.g., by overheating or the diffusion of Ga ions into the YBCO film during the FIB milling. The effective nanowire width that is available for conducting a super-current through the bridge is therefore reduced to $W_{eff} = W - W_d$, where $W$ is the geometrical wire width. The width of the damaged part of the nanowire exceeds the effective width by a significant amount for the thinnest wires. The damage turns the nanowire edges into insulating regions that do not influence the electrical measurements. Hence, these regions do not contribute either to the normal-state conductance or to the critical current at cryogenic temperatures. Together with the amorphous YBCO layer on top of the nanowire, the insulating YBCO regions at the edges protect the ultra-thin superconducting region at the center of the nanowire from diffusion of oxygen out of the film and degradation in moist air. We therefore refer to the effective rather than the geometrical widths of the fabricated nanowires below.

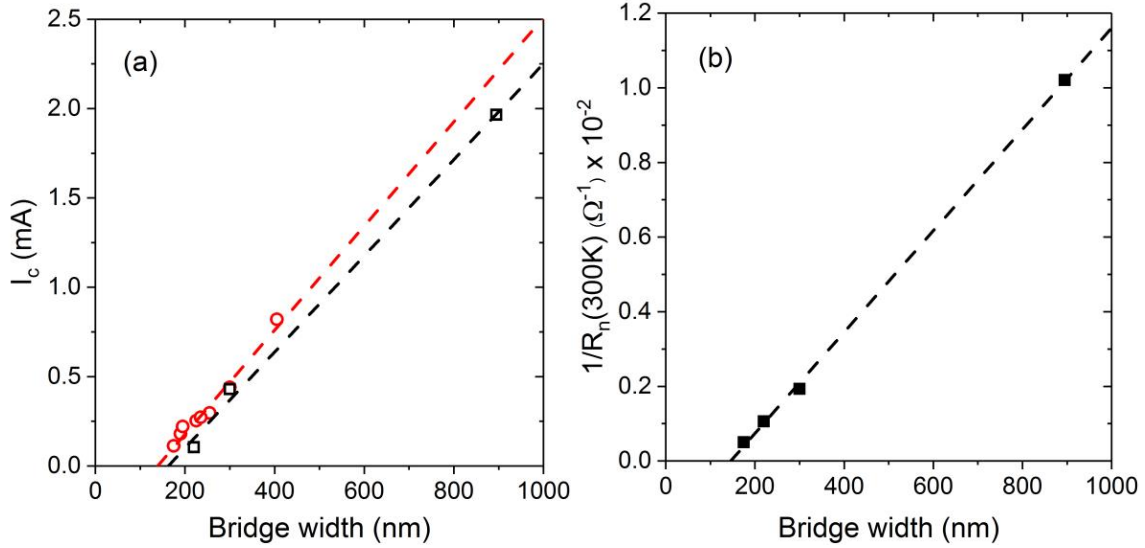

FIG. 3. (a) Dependences of critical current on nanowire width for 7 u.c. thick YBCO nanowires at $T = 4.2$ K (red hollow circles) and 20 u.c thick YBCO nanowires at $T = 78$ K (black hollow squares), and (b) dependence of room-temperature conductance of 20 u.c. thick YBCO nanowires on nanowire width (squares). The dashed lines are linear fits to the critical current and conductance data.

In order to calculate the critical and switching current densities in the ultra-thin YBCO films, the thickness of a non-superconducting YBCO layer at the YBCO/SrTiO$_3$ interface, which does not contribute to the super-current, has to be taken into account. The thickness of the non-superconducting YBCO layer was obtained by comparing the critical current densities in 10 u.c. and 4 u.c. thick bridges N6 and N10 of 150 nm and 5 μm width, respectively, measured at T = 10 K. The critical currents of 4 u.c. and 10 u.c. bridges were 5.31 mA and 1.50 mA, respectively. From the equation $[I_c(4\ \text{u.c.})/W(4\ \text{u.c.}-d_d)]=[I_c(10\ \text{u.c.})/W \cdot (10\ \text{u.c.}-d_d)]$, the thickness of the non-superconducting layer at the YBCO/SrTiO$_3$ interface can be calculated to be $d_d = 3.3$ u.c. This value is in a good agreement with our previous findings, where we observed that the first three YBCO layers on (001) SrTiO$_3$ substrates, i.e., 3 u.c. in thickness, contribute negligibly to the supercurrent through a film at $T = 78$ K and the critical current density is restored to the bulk value starting from the fifth layer [22]. We therefore normalize all of the nanowire current densities to the effective film thickness $d_{eff} = d - 3.3$ u.c below. The measured and effective widths and the length of the bridges and nanowires used in this work are listed in the Table I. The effective critical current density $J_{ceff}$ was determined by dividing the measured critical current by the effective nanowire width and thickness. For the 5-µm-wide and 20 u.c thick reference microbridge N1, the effective critical current density is inferred to be 11.6 MA/cm² at 78 K. However, for nanowires N2 and N3 of the same thickness but with widths of 750 and 155 nm, i.e., after FIB patterning, higher effective critical current densities of 13.6 MA/cm² and 14.2 MA/cm², respectively, are measured. The higher $J_{ceff}$ values can be attributed to a lower ratio of wire width, $W$, to the Pearl length $\Lambda = 2\lambda_{ab}^2/d_{eff}$, where $\lambda_{ab}$ is the magnetic penetration depth along the *ab*-plane. We calculate the $\Lambda(78K) = 6.4$ µm for N1-N5 samples using the effective film thickness $d_{eff} = 16.7$ u.c. and the estimated value of the magnetic penetration depth $\lambda_{ab} = 250$ nm at $T = 78$ K [24,25]. The value of $W/\Lambda = 0.78$ for the 5-µm-wide microbridge N1 decreases to $W_{eff}/\Lambda = 0.024$ for the nanowire N3 with the effective width of 155 nm resulting in more uniform supercurrent distribution across the nanowire and, hence, in the higher critical current density [26]. A further reduction in the nanowire effective width to 75 nm resulted in a decrease in the effective critical current density to 7.38 MA/cm² at $T = 78$ K. For the nanowire N5 with the effective width of 30 nm, no super-current could be measured at $T = 78$ K. The decrease in the critical current density correlates with the broadening of the superconducting transition width shown in Fig. 2.

TABLE 1. Parameters of microbridges and nanowires.

| Sample | Thickness (u.c.) | Effective thickness (u.c.) | Width (nm) | Effective width (nm) | Length (µm) | Critical current (mA) | Effective critical current density (MA/cm²) |
|---|---|---|---|---|---|---|---|
| N1 | 20 | 16.7 | 5000 | - | 7 | 11.3 at 78 K | 11.6 at 78 K |
| N2 | 20 | 16.7 | 895 | 750 | 1 | 1.99 at 78 K | 13.6 at 78 K |
| N3 | 20 | 16.7 | 300 | 155 | 1 | 0.430 at 78 K<br>3.20 at 8 K | 14.2 at 78 K<br>106 at 8 K |
| N4 | 20 | 16.7 | 220 | 75 | 1 | 0.108 at 78 K<br>0.886 at 8 K | 7.38 at 78 K<br>60.5 at 8 K |
| N5 | 20 | 16.7 | 175 | 30 | 1 | - | - |
| N6 | 10 | 6.7 | 300 | 150 | 1 | 1.50 at 8 K | 128 at 8 K |
| N7 | 10 | 6.7 | 350 | 200 | 1 | 1.68 at 8 K | 107 at 8 K |
| N8 | 7 | 3.7 | 195 | 55 | 2 | 0.220 at 4.2 K | 92.5 at 4.2 K |
| N9 | 7 | 3.7 | 400 | 260 | 2 | 0.820 at 4.2 K | 72.9 at 4.2 K |
| N10 | 4 | 0.7 | 5000 | - | 7 | 5.32 at 8 K | 130 at 8 K |

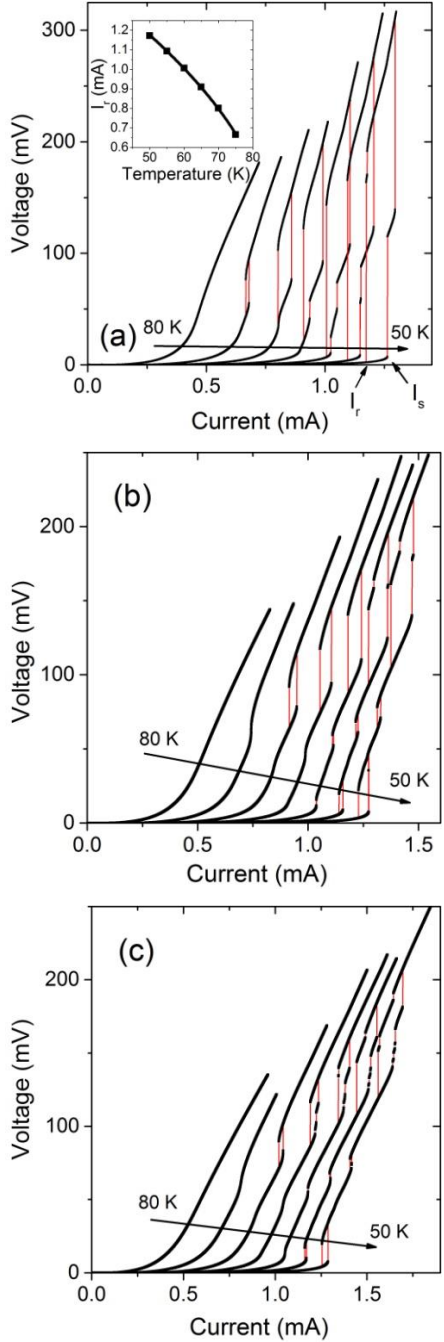

FIG. 4. *IV* curves measured from 1-μm-long, 150-nm-wide and 10 u.c. thick ($d_{eff} = 7$ u.c.) YBCO nanowires (a) without, (b) with 1 kOhm and (c) with 510 Ohm parallel connected resistors at $T = 50$-$80$ K with 5K steps. The red lines indicate voltage switching. Inset: temperature dependence of the retrapping current (squares) and fit using the hotspot model (solid line).

We associate this deterioration in the superconducting properties with overheating of the nanowires during FIB milling, resulting in a decrease in oxygen concentration and an increase in CuO-chain disorder.

## IV. HOTSPOT-ASSISTED VOLTAGE SWITCHING IN YBCO NANOWIRES

We first investigated the current-voltage characteristics of two 10 u.c. thick nanowires N6 and N7 with effective widths of 150 and 200 nm, for which the critical temperature and transition width were unchanged after FIB milling. The chosen nanowires have uniform super-current distributions over their cross sections due to the small values of $W_{eff}/\Lambda \leq 0.04$ in the temperature range $4 - 90$ K. The $\Lambda(0)$ was calculated as 5.0 μm for N6 and N7 nanowires using the effective film thickness $d_{eff} = 6.7$ u.c. and the zero-temperature value of the magnetic penetration depth $\lambda_{ab}(0) = 140$ nm [24]. In order to measure *IV* curves at temperatures below 50 K, a $0.5 - 1$ kΩ resistor was connected in parallel to the current-biased nanowires to avoid damage to them due to overheating after voltage switching.

Fig. 4a shows the *IV* characteristics of a 150-nm-wide and 10 u.c. thick nanowire N6 in the $50 - 80$ K temperature range. For $T \leq 75$ K, a flux-flow behavior is observed [27], followed by multiple hysteretic voltage switching at bias currents above the critical current. The differential nanowire resistance after voltage switching, $R_d = dV/dI \approx 600$-$1000$ Ω, was larger than the nanowire resistance at the onset temperature of the superconducting transition $R_n \approx 450$ Ω. As a next step a resistor was connected in parallel to the nanowire to evaluate an influence of the heating effects on the nanowire *IV* curve. Corresponding *IV* curves are shown for a 1 kΩ resistor in Fig. 4b and for a 510 Ω resistor in Fig. 4c. Both voltage switching amplitude and hysteresis were reduced by the parallel resistor, while the differential resistance remained the same. Using the 510 Ω parallel resistor, we found that the number of discernable resistive branches at $T = 50$ K increased from three for a bridge without a parallel resistor to eight for increasing the bias current.

Evaluating the influence of the parallel resistor on the nanowire *IV* curve, we conclude that the hysteresis in the *IV* curves is most likely associated with hotspots occurring along the nanowire. Within the framework of the hotspot model, the hysteresis in the *IV* curve appears when the switching current $I_s$ is higher than the minimum current $I_r$ required to sustain a normal-state domain by Joule self-heating. The $I_r$ value is given by

$$I_r \approx (2\alpha W^2 dT_c/\rho_n)^{1/2}(1-T/T_c)^{1/2} \qquad (1)$$

for bridges that are much longer than the thermal healing length $\eta = (kd/\alpha)^{1/2}$ or $I_r \approx (2\kappa d^2 T_c/\rho_n)^{1/2}(1-T/T_c)^{1/2}$ for bridges that are much shorter than the thermal healing length $\eta$ [16]. Here, $k$ is the thermal conductivity of YBCO, $\alpha$ is the heat transfer coefficient of the YBCO/SrTiO$_3$ interface, and $\rho_n$ is the normal-state resistivity. To choose relevant expression for the $I_r$ current we evaluate the nanowire length $L$ relative to the thermal healing length. For the 10 u.c. thick nanowire N6, we estimate the thermal healing length as $\eta \approx 80$ nm, using $\alpha = 850$ W/cm$^2$ K reported in [28] and $k \approx 0.05$ W/cm K derived from the electrical resistivity using the Wiedemann-Franz law. The calculated thermal healing length is much shorter than the nanowire length $L = 1$ μm. Therefore, we use the equation (1) to compare our experimental data with the hotspot model.

The consistency of our experimental data with the hotspot model was verified by fitting the temperature dependence of the retrapping current $I_r$, which is shown in the inset in Fig. 4a, to Eq. (1) using the zero-temperature retrapping current $(2\alpha W^2 dT_c/\rho_n)^{1/2}$ and the critical temperature as fitting parameters. The resulting curve, with fitting parameters $(2\alpha W^2 dT_c/\rho_n)^{1/2} = 1.80 \pm 0.01$ mA and $T_c = 87.0 \pm 0.2$ K, is shown using a solid line in the inset in Fig. 4a. The theoretical curve fits the experimental data very well. The fitting parameters are in good agreement with the measured critical temperature $T_c(10u.c.) = 86$ K and suggest a heat transfer coefficient $\alpha = 3400$ W/cm$^2$ K for the YBCO/SrTiO$_3$ interface, similar to values reported earlier [28].

Deeper insight into the origin of the voltage switching was obtained by measuring switching currents $I_s$ for 150 nm and 200 nm wide nanowires N6 and N7 in the $T = 8 - 80$ K temperature range, calculating the corresponding critical current densities $J_s$ and comparing them with relevant theoretical models. The switching current $I_s$ was defined as the current at which the first switching event is observed. At temperatures above 70 K, switching current values were determined from the positions of the first $dV/dI$ maxima.

The temperature dependences of the effective switching current densities $J_s(T) = I_s(T)/W_{eff}d_{eff}$ for nanowires N6 and N7 are shown in Fig. 5. At a temperature $T$ of 10 K, the effective switching current densities of these nanowires were $J_s(W_{eff}=150$ nm$) = 145$ MA/cm$^2$ and $J_s(W_{eff}=200$ nm$) = 136$ MA/cm$^2$. We compare our measurements to calculations of the current density $J_s(AL) = (\Phi_0/4\mu_o\lambda_{ab}^2)(W\xi_{ab})^{-1/2}$ required to suppress the edge barrier of a superconducting bridge within the Aslamazov-Larkin (AL) model, where $\Phi_0$ is the magnetic flux quantum, $\mu_o$ is the vacuum permeability, and $\xi_{ab}$ is the coherence length in the *ab*-plane [29,30]. The $J_s(AL)$ was calculated using estimated values for the magnetic penetration depth $\lambda_{ab}(10K) = 140$ nm [24,25] and the coherence length $\xi_{ab}(10K) = 1.3$ nm [31]. The calculated values $J_s(W=150$ nm$) = 150$ MA/cm$^2$ and $J_s(W=200$ nm$) = 136$ MA/cm$^2$ are in good agreement with the measured switching current densities.

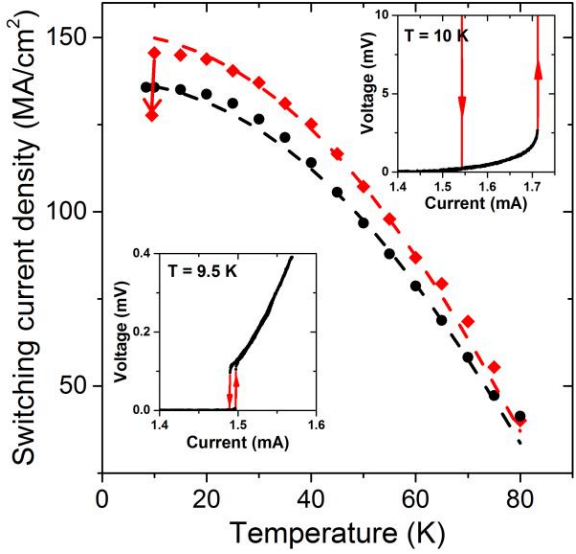

FIG. 5. Temperature dependences of the effective switching current density for 150-nm-wide (red diamond) and 200-nm-wide (black circle) nanowires N6 and N7, respectively. The dashed lines are switching current densities calculated according to AL model. Insets: *IV* curves of the 150-nm-wide nanowire N6 at $T = 9.5$ K and $T = 10$ K. The red lines in the insets indicate voltage switching.

The validity of the AL model extends over the entire temperature range up to the critical temperature. However, at present the $\lambda(T)$ and $\xi(T)$ dependencies are unknown for ultra-thin YBCO films. We therefore compared our measured data with $J_s(T)$ calculations, in which we assume a temperature-dependent magnetic penetration depth and coherence length in the form $\lambda_{ab}(T),\xi_{ab}(T)= \lambda_{ab}(0),\xi_{ab}(0)\ (1-(T/T_c)^a)^b$. The best agreement with our experimental data was found for exponents $a = 2$ and $b= -0.5$ for the magnetic penetration depth and $a = 4$ and $b = -0.5$ for the coherence length using $\lambda_{ab}(0) = 140$ nm and $\xi_{ab}(0) = 1.3$ nm [31]. The calculated curves, which are shown in Fig. 5 as dashed lines, are in good agreement with our experimental data.

The exponents *a* and *b* found for the $\lambda_{ab}(T)$ dependence in YBCO nanowires are close to those earlier reported for YBCO thin films [32,33] and the same as in the expression $\lambda(T)= \lambda(0)/(1-(T/T_c)^2)^{1/2}$ which produces good fitting to the microscopic BCS calculation for s-wave superconductors [34]. Since the $\lambda(T)$ in YBCO depends on a impurity scattering, the result close to the BCS calculations for s-wave superconductors rather than for d-wave superconductors, where $a = 4/3$, could be due to the scattering at the nanobridge edges [34].

Remarkably, that, as it is seen from Fig. 4a, the voltage switching occurs only for those kinks in the *IV* curve, corresponding to the current $I_s(AL) = J_s(AL)Wd$, where $I_s(AL)$ is larger than the minimum hotspot current $I_r$ given by Eq. (1). Thus, the first voltage switching, observed in the nanowires N6 and N7, can be explained by the hotspot-assisted suppression of the edge barrier by the transport current.

## V. PHASE-SLIP LINES IN YBCO NANOWIRES

At low temperatures, for nanowires with small cross-sections we observe direct voltage switching from the superconducting to the resistive state at bias currents below predictions of the AL model, as indicated by an arrow in Fig. 5 for the nanowire N6 dataset. The *IV* curves demonstrating the first switching event for the nanowire N6 at temperatures of 9.5 K and 10 K are shown in the insets in the Fig. 5. At $T = 10$ K the voltage switching occurs from the flux-flow state, while at $T = 9.5$ K the nanowire switches to the resistive state directly from the superconducting state. Both the switching current and the voltage switching amplitude are significantly lower for the direct voltage switching at $T = 9.5$ K. Furthermore, the differential resistance after voltage switching in such nanowires is significantly lower than the normal-state nanowire resistance. These findings indicate that direct voltage switching is not induced by a vortex flux flow instability, but originates from a different mechanism.

We therefore recorded the *IV* characteristics of several nanowires of the effective thickness $d_{eff} = 3.7–16.7$ u.c. and widths $W_{eff} = 30\text{-}260$ nm having cross-sections $W_{eff}d_{eff} \leq 1500$ nm$^2$ that corresponds to $W_{eff}/\Lambda \leq 0.037$. The $\Lambda(0)$ was calculated for these nanowires as $2.0 - 9.0$ μm using the zero-temperature value of the magnetic penetration depth $\lambda_{ab}(0) = 140$ nm [24]. The direct voltage switching was observed for all such nanowires, with amplitudes ranging from tens of μV for the widest nanowires to a few mV for nanowires narrower than 100 nm. *IV* curves of nanowires with larger cross-sections demonstrated flux-flow behavior, with a resistive slope emerging from the superconducting branch.

Fig. 6 shows representative *IV* curves for 2-μm-long 3.7 u.c. thick YBCO nanowires N8 and N9 with effective widths of 55 and 260 nm. For the 55-nm-wide nanowire N8, the largest voltage switching amplitude of 3 mV was observed, as shown in Fig. 6a. The corresponding *IV* curve consists of a superconducting branch and three resistive branches with differential resistances $R_d = dV/dI$ of 73, 148 and 250 Ω, which are separated by voltage switching events. Here we notice several features that are characteristic to phase-slip process in superconducting bridges: direct voltage switching from the superconducting to the resistive state; the differential resistance of each resistive branch is proportional to the branch number (counting from the lowest *I(V)* values up); the resistive branches have approximately the same excess current $I_{ex}$ [27]. The excess current $I_{ex}$ was determined from the intersection of the resistive branch slope with the current axis. Taking into account the large width of our nanowires $W >> \xi$, we assume that the appearance of phase-slip lines (PSLs) is responsible for direct voltage switching. Applying the resistance of a single phase-slip line $R_{PSL} = 73$ Ω, obtained from a linear fit to the first resistive branch, and the phase-slip center model (PSC) [27], we calculate the current-voltage dependence separately for the n$^{th}$ resistive branch as $V_n = nR_{PSL}(I - I_{ex})$, as shown in Fig. 6a using an orange line. Good agreement between the experimental and calculated *IV* curves indicates that the nanowire properties are uniform along the nanowire N8 and the heating effects are small despite of the large voltage switching

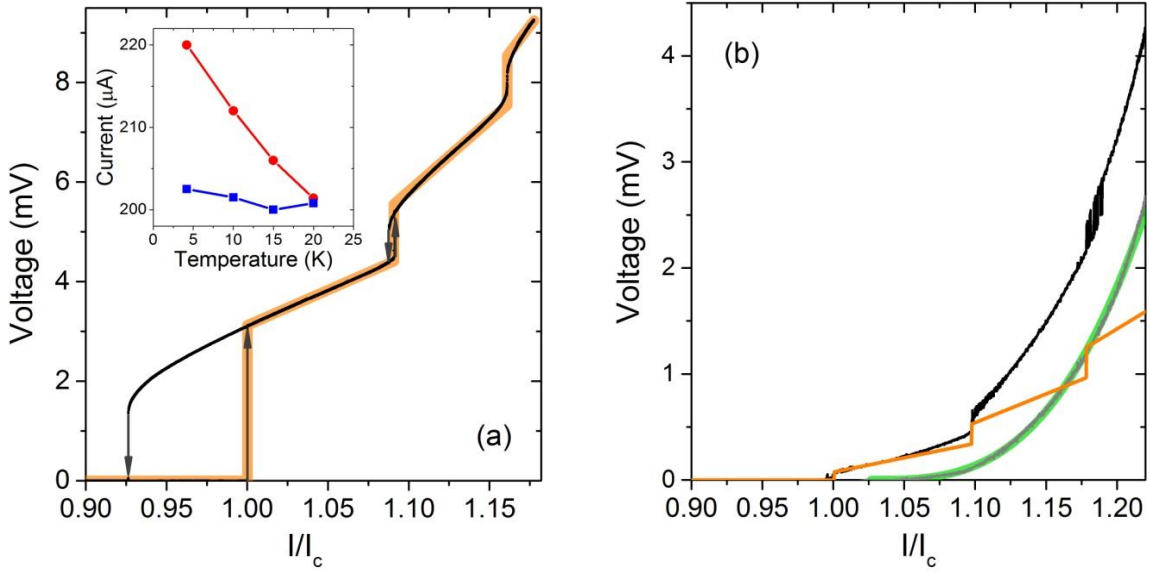

FIG. 6. (a) *IV* curve of a 55-nm-wide YBCO nanowire N8 at *T* = 5 K (black line) and *IV* curve calculated using the PSC model (orange line). Inset: temperature dependences of switching (red circle) and retrapping (blue square) currents for the nanowire N8. (b) *IV* curve of a 260-nm-wide YBCO nanowire N9 at *T* = 4.2 K (black line), the calculated PSL contribution (orange line), the extracted Abrikosov vortex contribution (gray line), and the fitting of the Abrikosov vortex contribution with Eq. (2) (green line).

amplitude.

The *IV* curve of the 260-nm-wide nanowire N9, for which a lower edge barrier is expected, shows a pronounced nonlinear behavior, in addition to direct voltage switching, as shown in Fig. 6b. Assuming that this *IV* curve is a sum of a PSL and some nonlinear components, we calculated the nonlinear component by subtracting the PSL contribution, as shown in Fig. 6b by an orange line, from the experimental data. The resistance of the single PSL $R_{PSL}$ = 3.3 Ω was obtained from the linear fit of the *IV* curve close to the first switching event at a current $I_s$ of 0.82 mA. The resulting nonlinear component is shown in Fig. 6b using a gray line. The calculated nonlinear contribution can be explained in terms of Abrikosov vortex motion, which follows a power-law dependence of the form

$$V \sim (I-I_{cAV})^\mu , \quad (2),$$

where $I_{cAV}$ is the critical current for Abrikosov vortex motion [35]. A power-law exponent was defined as $\mu = 3.02 \pm 0.02$ from a linear fit of the nonlinear contribution on a log-log scale, which agrees well with the expected value ($\mu = 3$) for thin-film superconductors in zero magnetic field [35]. We then obtained the critical current for Abrikosov vortex motion $I_{cAV}$ to be $0.842 \pm 0.002$ mA by fitting the nonlinear contribution to Eq. (2). The result of the nonlinear fit is shown in Fig. 6b by a green line and coincides with the nonlinear contribution over the entire current range. A good fit of the equation (2) to the experimental data indicates that the motion of the Abrikosov vortices is not influenced by the phase-slip lines and these processes can be considered as independent. At temperatures above 10 K, the critical current of the Abrikosov vortices in the 260-nm-wide nanowire N9 becomes smaller than the critical current of PSL, the direct voltage switching disappears, and the resistive slope emerges from the superconducting branch. Thus, we can explain the nonlinear *IV* curve of the 260-nm-wide nanowire N9 by the coexistence of PSLs with the motion of Abrikosov vortices.

Similar PSL dynamics was previously observed for wide Sn and InSn bridges at temperatures close to their critical temperature [6]. The presence of PSL dynamics in wide $W \gg \xi$ bridges was explained by the motion of kinematic vortices crossing the bridges [6,36]. Assuming that only one kinematic vortex is present in a PSL at a time, we estimate the average kinematic vortex velocity for sub-100-nm-wide YBCO nanowires to be $v_{kv} = W/\tau = WV_s/(2e/h) = 5 \cdot 10^4$ m/s, where $\tau$ is the vortex crossing time, $V_s$ denotes the voltage switching amplitude, $h$ is the Plank constant and $e$ is the electron charge. This value is two orders of magnitude larger than the measured velocities of Abrikosov vortices in YBCO films [13,14] and is similar to kinematic vortex velocities in low-$T_c$ superconductors ($v_{kv} \approx 10^5$ m/s [6]). The retrapping current $I_r$ does not follow a $(1-T/T_c)^{1/2}$ dependence, but is practically temperature-independent, as shown in the inset to Fig. 6a, which provides evidence that the hysteresis in these *IV*-curves does not originate from the hotspot effect, but rather from the intrinsic properties of the phase-slip process [37].

Finally, we calculated the penetration depth of the electric field $\Lambda_E = R_{PSL}Wd/2\rho_n = 235\pm110$ nm [6,7]. In our calculations, we used the normal-state resistivity $\rho_n = 4$ μΩ·cm at $T = 5$ K, assuming a linear temperature dependence $\rho_n(T)$ [15]. Alternatively, the penetration depth $\Lambda_E$ was determined from the voltage switching currents. For weakly interacting PSLs, the corresponding bias current values are given by $I_{s,n} = I_c (1+\exp(-L/2n\Lambda_E))$, where $L$ is the nanowire length [38]. By evaluating the $n = 2$ and $n = 3$ voltage switching events in Fig. 6, we find the electrical field penetration depth to be $\Lambda_E = 200\pm14$ nm, which is similar to the $\Lambda_E$ value obtained from the PSL resistance.

Based on the above, we explain the direct voltage switching observed at low temperatures for the YBCO nanowires with small cross section by the appearance of the phase-slip lines in the nanowire.

## VI. CONCLUSIONS

We have observed two types of current-induced voltage switching in the *IV* curves of ultra-thin YBCO nanowires. Voltage switching in the flux-flow *IV* curves occurs due to the hotspot-assisted suppression of the edge barrier by the transport current. Here, the bias current values at which hotspot-assisted voltage switching occurs is in good quantitative agreement with the predictions of the Aslamazov-Larkin model for wide superconducting bridges. The direct voltage switching of nanowires with small cross-sections ($Wd \leq 1500$ nm$^2$) is attributed to the appearance of phase-slip lines in the YBCO nanowire.

## ACKNOWLEDGMENTS

The authors would like to thank M. Kupriyanov and Y. Divin for valuable discussions and I. Gundareva for help with experiments. This work was partially supported by ER-C project C-088. C. Schuck acknowledges financial support from the Ministry of Culture and Science NRW.

———————————————————————